\documentstyle[prb,aps,epsfig]{revtex}
\begin{document}
\draft
\title{Photovoltage  Effects in Photoemission from Thin GaAs Layers}
\author{G. A. Mulhollan\cite{Mulhollan}, A. V. Subashiev\cite{Subashiev}, 
J. E. Clendenin, E. L. Garwin, R. E. Kirby, \\ and T. Maruyama} 
\address{Stanford Linear Accelerator Center, Stanford, California
94309}
\author{R. Prepost}
\address{Department of Physics, University of Wisconsin, Madison, Wisconsin
53706}
\maketitle
\begin{abstract}
\renewcommand{\baselinestretch}{1.50}
\small\normalsize
A set of thin GaAs $p$-type negative electron affinity (NEA) photocathodes
have been used to measure the yield of photoemitted electrons at high
intensity excitation.
The active layer thickness is 100 nm and the $p$-type doping ranges from
5$\times$10$^{18}$~cm$^{-3}$ to 5$\times$10$^{19}$~cm$^{-3}$ for a set
of four samples. 
The results show that the surface escape probability is a linear function
of the NEA energy. 
The surface photovoltage effect on photoemission is found to diminish 
to zero at a doping level of 5$\times$10$^{19}$~cm$^{-3}$. 
The experimental results are shown to be in good agreement
with calculations using a charge limit model based on surface photovoltage
kinetics that assume a constant  electron energy relaxation 
rate in the band bending region. 
\end{abstract}
\pacs{29.25.Bx, 73.50.Pz, 73.61.Ey, 79.60.-i, 85.60.H} 
\pagebreak
\section {Introduction.}
The negative electron affinity (NEA) state of activated (100) GaAs
has been widely studied and used for a variety of applications.
While the qualitative features of the electron photoemission are
known, a full quantitative understanding is lacking
\cite{herman,rev}.
The co-adsorption of Cs and O (or F) on an atomically clean surface 
of $p$-type GaAs is known to result in a significant lowering of
the GaAs electron affinity as well as a shift of the surface
Fermi level deep into the band gap with a formation
of the band bending region (BBR) near the surface. 
The band-bending is an important contribution to the lowering
of the electron affinity.  The emitted electron energy distribution
depends on the density of states, the doping level
and the electron kinetics in the BBR, leading to differing
interpretations of the electron
energy distribution (EDC) curves \cite{baum,Mam,Ter1,Nolle}.
While the activated surface develops a surface barrier for the electrons,
the structure and transparency of the barrier is not well understood.

Information on the properties
of the NEA surface is obtained with photoemission
studies at high power excitation \cite{Tang,Woods,Rez}.
When the photocathode is excited with high densities of light 
near the band gap,
the total photoemitted charge is not proportional to the light intensity.
This phenomenon was first observed using bulk GaAs \cite{Woods}, and
can be described as a surface charge limit (SCL).
Several experimental studies have been made on strained GaAs\cite{Tang},
superlattice structures \cite{Nakani}, and thin unstrained GaAs \cite{Jarosh}.
In \cite{Spicer} the  SCL was attributed to the
photovoltage building up in the band bending region.
The photoelectrons captured in
the BBR produce an opposing field that flattens the bands
and reduces NEA, which can be described as the effect of the surface
photovoltage (SPV). 
It was  determined that SPV formation
and its relaxation is mainly controlled by the restoring
current of the holes \cite{Hecht} and is very sensitive
to the transparency of the barrier for the holes formed
by the BBR \cite{Molod}. The value of the surface photovoltage
at several temperatures for a (100) GaAs surface has been 
measured in studies of core-level photoemission
spectra under synchrotron radiation excitation \cite{Molod}
and in studies of the EDC at high excitation levels
\cite{Jarosh}.

The problem of the SCL is important for operation of
polarized electron photocathodes at the Stanford Linear Collider
(SLC)\cite{Alley} 
and even more so for the next generation linear
colliders such as the Next Linear Collider (NLC)\cite{NLC}. 
The present NLC design requires a train of 95 microbunches 
spaced 2.8 ns apart. Each microbunch is required to have 
$2.8\times 10^{10}$ electrons with a peak current of 4.5 A,
totaling $2.7\times 10^{12}$ electrons in a 266 ns train. The 
required total charge is two orders of magnitude higher than
the SLC case. 

In the present paper, we report on the first
detailed experimental studies and theoretical analysis
for a set of 100 nm-thick GaAs samples with various levels 
of uniform doping.
For thin-layer cathodes the time of electron
extraction from the active layer to the BBR is much smaller
than the bulk lifetime and the diffusion time,
eliminating the influence of the recombination
effects in the layer. The optical properties of
unstrained GaAs are fairly well known, 
allowing reliable estimation of the surface
escape probability from the experimental data.
Using long-pulse excitation and pump-probe measurements, 
the kinetics of photovoltage buildup and electron emission are
investigated.

\section{The Charge Limit Model}
\subsection{Surface Escape Probability}
Fig. 1 shows schematically the GaAs band structure near the
surface activated by Cs(O/F) deposition and the electron potential
near the surface.
When the cathode is illuminated with a pulsed laser with photon energy slightly
above the band gap energy, the photoexcited electrons in the conduction
band are rapidly thermalized and captured in the BBR potential well from which 
they tunnel into vacuum through the surface barrier.
For a thin unstrained GaAs layer ($\alpha d \ll 1 $,\ 
$\alpha L \ll 1 $, where $\alpha $ is the optical absorption
coefficient, $d$ is the active layer thickness, $L$ is the diffusion length)
all the electrons optically excited in
the GaAs layer  near the absorption edge are captured
in the BBR region forming the flow of the electrons
to the surface, $j_{\rm el} = q\alpha d (1-R)J$,
where $J$ is the light excitation intensity and $R$ is the surface 
optical reflection coefficient. A part of this flow is emitted into vacuum,
while the rest contributes
to the surface recombination current, so that
the quantum yield Y is\cite{rev}:
\begin{equation}
Y = \alpha d (1-R) B_N\ . 
\label{eq:Y}
\end{equation}
Here $B_N$ is the surface escape probability. The value of $B_N$ 
depends on the details of the competition between 
electron recombination in the BBR and emission
into vacuum through the surface barrier.

To estimate local values of $B_N$ on the activated surface
we use the results of near band gap low-intensity measurements of the quantum
yield for a small excitation spot. Taking  $d=10^{-5}$ cm,
$(1-R)$=0.68 and $\alpha = 5 \times 10^3$ cm$^{-1}$
\cite{Casey}, we obtain $B_N \approx 0.15$ for    
a typical quantum yield of a thin GaAs layer of $Y$ = 0.5\%.
Thus, even at the low-intensity excitation 
only $1/7$th of the electron flow to the
surface is photoemitted into vacuum, and 
the majority recombines in the band bending region.

The quantum yield from NEA surfaces increases with  
bias field at the GaAs surface \cite{Howorth,Ter2}.
This effect is attributed traditionally to the image-force lowering 
of the surface barrier by the applied field (see Fig. 1a).
This barrier lowering, $\delta U$, can be evaluated from a
simple electrostatic consideration:
\begin{equation}
\delta U(F) = q\sqrt {{q F (\epsilon_s-1)} \over{4 \pi
\epsilon_0(\epsilon_s+1)}}, 
\label{Aten}
\end{equation}
where $q$ is the free electron charge,
$F$ is the external electric field, and $\epsilon_s$ is the relative 
permittivity of the semiconductor. The quantity $\delta U(F)$ is seen to be 
proportional to the square root of the bias field. 
More commonly, this effect is studied for surfaces having positive electron 
affinity or Schottky barriers, where the image force modifies the thermionic 
emission current. The emission current is then an exponential function 
of the barrier height, so that  $\log(Y)$ grows linearly 
with the square root of applied bias field. For NEA photocathodes,
both linear\cite{Ter2} and logarithmic\cite{Tang} dependences
of $Y$ on the square root of
the bias field have been observed. 
In Appendix A an expression for the surface escape probability
$B_N$ is derived. For the conditions of this experiment the
dependence of $B_N$ on the NEA value $\Delta$ is shown to be linear.
From these considerations it follows that the relative variation 
of the yield with the bias field F is given by:
\begin{equation}
\frac{Y(F)}{Y(0)} = 1 + \frac{\delta U(F)}{\Delta}. \label{bias}
\end{equation}

\subsection{The Photovoltage}
Under intense optical excitation 
the electron flow to the surface starts to compensate the positive charge of
the donor-like states that provide band bending at the surface.
This results in a flattening of the energy bands and a lifting
of the bottom of the conduction band at the surface (see the dashed curve in
Fig. 1b). 
One can assume that the properties of the resulting 
effective surface barrier in the several monolayer-thick activation 
layer are not influenced by this relatively small shift. One can
further assume that the electron diffusion to the band bending
region does not change since it remains faster than the
surface processes. The change in quantum yield can then be evaluated
if one replaces the NEA value $\Delta$ by
$\Delta - U(J)$, where $U(J)$ is the up-shifting
of the conduction band due to the photovoltage. Taking into account
the bias field effect, we obtain
\begin{equation}
\frac{Y(J)}{Y_0} = 1 - {{ U(J)} \over{\tilde \Delta}},
\label{PhVo}
\end{equation}
where $\tilde\Delta =\Delta+\delta U$ is the
NEA value recalculated for a given bias field, $\delta U$ is the 
potential barrier lowering resulting from an external bias field, 
and $Y_0$ is the quantum yield for non charge-limited photoemission. 

\subsection{The Restoring Currents}
When the excited electrons are captured in the BBR, 
the majority of electrons recombine at the surface with the holes coming 
from the
valence band. There are several mechanisms for electron
recombination (starting from the recombination in the band
bending region itself). However, for activated GaAs surfaces
the electron capture by surface localized states
appears to be the fastest process since the electrons drift
in the strong electric field to these capture centers, which have 
an attractive Coulomb potential.
Therefore the recombination rate of these electrons is limited
by the excess hole current to the surface, $j_p$.
In the stationary state where the excitation time is much longer than the
surface process time, the SPV value can be evaluated from the balance
of the surface current densities, $j_{el}=j_p$.

There are two major mechanisms providing the recombination hole current
to the surface through the band bending
region: thermionic emission and tunneling through
the barrier. The photovoltage dependence of the hole current
at low temperatures tracks with the tunneling contribution, while
at higher temperatures the hole current increases due to the thermal shift
of the hole energy distribution, which assists hole tunneling through the barrier. 
In the equilibrium state with no illumination, the excess hole
current equals zero at the equilibrium band bending.
Therefore, the resulting excess current $j_p(U)$ can be expressed
as\cite{Rho}:
\begin{equation}
j_p(U)=j_{p,0}[\exp(U/E_0)-1],
\label{eq:JhoU}
\end{equation}
where the energy $E_0$ for the band bending region is given by:
\begin{equation}
E_0=E_{00}\coth({E_{00}\over{kT}}),\ \  
E_{00}=\frac{\hbar}{2} \left[\frac{q^2N_a}{m_h
\epsilon_0\epsilon_s} \right]^{1/2}=
18.57\times\left[\frac{N_a}{m_h\epsilon_s}\right]^{1/2} {\rm meV},
\label{eq:Eoo}
\end{equation}
where $m_h$ is the hole mass, and 
$N_a$ is the acceptor concentration in the unit of
10$^{18}$ cm$^{-3}$. 
For the case of thermally assisted tunneling in metal-semiconductor junctions,
$j_{p,0}$ can be expressed as:
\begin{equation}
j_{p,0} = A^{**}\frac {T (E_{00} V)^{1/2}} {k \cosh(E_{00}/kT)}
\exp(-{V \over E_0}),
\label{eq:j0}
\end{equation}
where $k$ is the Boltzmann constant and $A^{**}$ is the
effective Richardson constant, $A^{**}$ =9.6 A K$^{-2}$ cm$^{-2}$,
for tunneling by light holes. 
It follows from
Eq. (\ref{eq:Eoo}) that for GaAs and a typical doping level of
5$\times$ 10$^{18}$ cm$^{-3}$, the value of $E_{00}$ can vary
from 17 meV to 40 meV depending on the relative contribution
of the light and heavy holes into the hole current. 
However, the light hole contribution to $j_p(U)$ is dominant
due to the mass dependence of the $j_{p,0}$ factor. 
The above expression for $j_{p,0}$ is in the context of a Schottky
barrier model with a metallic layer and thus may be different for 
cesiated NEA GaAs surfaces. 

\section{Experiment}

The samples for the present experiment were grown by
the Quantum Epitaxial Devices Corporation
\cite{QED} using molecular-beam-epitaxy (MBE).
The substrate material was (100) {\it n}-type (Si doped to
1$\times$10$^{18}$cm$^{-3}$) GaAs. Since heavy {\it p}-type doping is
necessary to achieve a NEA surface, a 0.1-$\mu$m-thick {\it p}-type
GaAs (Be doped to 5$\times$10$^{18}$cm$^{-3}$)
layer was first grown on the substrate, followed by a 1-$\mu$m-thick {\it p}-type
(Be doped to 5$\times$10$^{18}$cm$^{-3}$)
Al$_{0.3}$Ga$_{0.7}$As buffer layer. The 0.1-$\mu$m-thick {\it p}-type GaAs
active layer was then grown on this buffer.
Four samples were grown
with uniform doping concentrations of 5$\times$10$^{18}$cm$^{-3}$,
1$\times$10$^{19}$cm$^{-3}$, 2$\times$10$^{19}$cm$^{-3}$,
and 5$\times$10$^{19}$cm$^{-3}$ in the active layer respectively.
The Al$_{0.3}$Ga$_{0.7}$As intermediate layer serves as a potential barrier
to isolate the active GaAs layer from the substrate GaAs.
In order to preserve an atomically
clean surface the samples were anodized to form an oxide layer of
about 50 \AA~ on the GaAs surface\cite{Schwartz}.
The oxide layer was later removed as described below.
The relevant sample parameters are tabulated in Table I. 

The experiments were performed at the Gun Test Laboratory at SLAC
\cite{Alley}. The apparatus, which is a replica of the first few meters
of the SLAC injector beamline, consists of a 22.5 mm diameter photocathode
diode gun with a load-lock system for cathode loading/removal,
and an electron beamline terminating into a Faraday cup.
Prior to installation in the system, the sample was degreased in a boiling
solution of trichloroethane. After the protective oxide layer was removed in
ammonium hydroxide, the sample was rinsed in distilled water and methanol.
The load-lock system was essential for loading the samples into and
removing them from the gun under vacuum. This system avoided the cathode
contamination that normally accompanies the bake of the gun system, resulting
in reproducible cathode activations and high quantum yield.
The cathode activation to obtain an NEA surface consisted of heat cleaning
to 600 $^\circ$C for 1 hour, followed by 
application of cesium until the photoyield peaked, and then cesium and
nitrogen-trifluoride codeposition until the photoyield was again maximized.

Two different excitation sources were used for the two types of 
measurements. The first type of measurement used a flash-lamp 
pumped Ti:sapphire laser (Flash-Ti) producing a long laser pulse 
adjustable from 150-350 ns at 120 Hz with energy up to 130 $\mu$J.
This excitation mode was used to directly study the time evolution
of the photovoltage and the dependence of the photovoltage on the 
laser intensity.
The second type of measurement used 
two pulsed Ti:sapphire lasers pumped by a single frequency-doubled 
Nd:YAG laser (YAG-Ti) in a pump-pulse/probe-pulse combination.
In this excitation mode the pump-pulse was an intense short
pulse inducing a photovoltage in the photocathode followed by
a short high intensity probe-pulse, but not in the intensity regime to 
produce a photovoltage effect. The two YAG-Ti lasers each produced
a 2 ns pulse at 60 Hz with a pump laser energy of up to 100 $\mu$J.
In these measurements, 
the time separation between the pump and probe laser pulses was
varied. This technique allowed 
an independent measurement of the photovoltage parameters.
All laser wavelengths were 850 nm, and the spot size was set to 
about 20 mm so as to fill the exposed area of the cathode.  

An optically isolated nanoammeter, a beam position monitor
(BPM), a ceramic gap monitor, and a fast Faraday cup were used for beam 
intensity measurements. The cathode was biased at $-120$ kV and maintained
at a temperature of 0 $\pm$ 2$^\circ$ C.
The vacuum in the gun was maintained at about
1x10$^{-11}$ Torr by means of ion and nonevaporative getter pumping.

Two low-power CW diode lasers of 833 nm and 850 nm were used for 
the low power quantum
yield measurements. For an individual sample, the quantum yield 
typically varied by a factor of two over the surface,
but in some cases as high as a factor five. 
These variations were presumably caused by the inhomogeneous distribution of 
cesium on the surface. Similar variations were reported in Ref. \cite{Ter2}.
The quantum yield values averaged over the surface for the individual
samples are shown in Table 1.  

\section{Results}
The bias field dependence of the quantum yield was measured 
using the low power 833 nm diode laser.
Fig. 2 shows the experimentally observed variation of the quantum yield
with the bias field for sample 1a, sample 3, and two other photocathodes
not used for the present experiment. These two additional samples
were zinc-doped strained GaAs and carbon-doped 1 $\mu$m thick GaAs.
The observed dependence clearly favors
a linear rather than a logarithmic dependence of $Y$ on $\sqrt{F}$.
Following Eq. (\ref{bias}), the zero-field NEA value $\Delta$ can
be determined from the slope of the fitted curves, shown as a
solid line in Fig. 2, giving values of $\Delta$
in the range  $\Delta=$ 122 -- 138 meV. These NEA values
are, to a good approximation, the same for all well activated samples. 
As a comparison, the values extracted from the data of Ref.
\cite{Ter2} for different points on the surface of a thick GaAs
cathode are in the range of 100 -- 160 meV. Therefore the quality
of the activation in both cases is found to be similar.

The surface charge limit effect can be explored
if the cathode is excited with a high peak power laser using a pulse length
much longer than the characteristic time scale of the surface photovoltage
effect. This is the technique using the long-pulse Flash-Ti laser described
earlier. Figs. 3a, 3b, and 3c
show representative temporal profiles of the emission current
pulses measured using varying light pulse energies  
for (a) sample 1b, (b) sample 2a, and (c) sample 3, respectively. 
The observed photovoltage effect has several features. 
As seen in Fig. 3a, the electron emission current rises to a peak at
the start of the laser pulse and then decreases as
the photovoltage builds up. With time the electron
emission current reaches a steady state value as the photovoltage
saturates. With increasing laser energy, the photovoltage builds
up more quickly and the suppression of the emission current due to
the photovoltage is more pronounced. Fig. 3b and Fig. 3c show that the 
photovoltage effect decreases as the doping level increases.  

To understand the temporal profile of the emission current, the 
charge limit model described in Section II is used.
The time variation of the excess electronic charge $Q$ at the surface
can be written in terms of the capacitance per unit surface area  
and the restoring hole current as:
\begin{equation}
C\frac{dU}{qdt}= j_{el}-j_p(U) = q\alpha d (1-R)J - j_p(U).
\label{eq:Kin1}
\end{equation}
Within the depletion region approximation \cite{Rho} for the BBR
region with width $w$, the capacitance is given by:
\begin{equation}
C=\epsilon_0\epsilon_s/w = 
\sqrt{\frac {q^2 \epsilon_s \epsilon_0 N_a}{2 (V-U)}},
\label{eq:Cap}
\end{equation}
where $V$ is the initial band bending energy (see Fig. 1).
Here we  assume that the emission current is much smaller
than the recombination current since the surface escape probability 
$B_N \approx$ 0.15.
Integration of Eq. (\ref{eq:Kin1}) using Eq. (\ref{eq:JhoU})
gives for the normalized yield:
\begin{equation}
\frac{Y(J)}{Y_0} = 1 + \frac{E_0}{\tilde \Delta} \times \ln \left[
\frac{1+ qJ/\tilde j_{p,0}\exp[-(1+qJ/{\tilde j_{p,0}})t/\tau]}
{(1+qJ/{\tilde j_{p,0}})}\right],
\label{eq:Kin3}
\end{equation}
where ${\tilde j_{p,0}}$ = $j_{p,0}/\alpha d (1-R)$ and $\tau = E_0C/qj_{p,0}$.
The quantum yield decreases with a characteristic saturation 
time constant $\tau^\prime$, where
\begin{equation}
\tau^\prime = \frac{\tau}{(1+qJ/\tilde j_{p,0})}, \label{eq:tau}
\end{equation}
and reaches the steady state value:
\begin{equation}
Y(J)/Y_0 = 1-\frac{E_0}{\tilde \Delta} \ln(1+{qJ\over\tilde
j_{p,0} })\ . \label{eq:Yfin}
\end{equation}

Figs. 4a, 4b, and 4c show
the measured  dependencies of the steady state emission current
on the light excitation intensity for samples 1a and 1b (Fig. 4a), 
samples 2a and 2b (Fig. 4b), and sample 3 (Fig. 4c) respectively,
measured for a broad range of excitation intensities. 
The solid lines in Fig. 4 are fits for $J\cdot Y(J)$ using Eq. (\ref{eq:Yfin}).
The SCL effect appears as the deviation of the emission current
from a linear dependence on the excitation light intensity
and is more pronounced in the lower doped samples (samples 1a and 1b).
The SCL effects are not observed in the samples
with {\it p} = 2$\times$10$^{19}$cm$^{-3}$ (sample 3),
and 5$\times$10$^{19}$cm$^{-3}$ (sample 4, not shown in Fig. 4).

Figs. 5a and 5b show
the experimentally observed variation of the saturation time $\tau^\prime$
as a function of the excitation intensity
for (a) samples 1a, 1b, and (b) samples 2a, 2b.
The saturation time shortens as the excitation intensity is increased,
consistent with Eq. (\ref{eq:tau}).
The solid lines in Fig. 5 represent the results of a fit to 
Eq. (\ref{eq:tau}).  

The decay of the surface photovoltage can also be explored 
using the pump-probe technique described earlier. 
The short intense pump laser pulse induces a photovoltage which 
subsequently decays in time according to:
\begin{equation}
C\frac{dU}{qdt} = -j_p(U).\label{eq:Udt}
\end{equation}
The solution of Eq. (\ref{eq:Udt}) (neglecting the weak variation of 
the BBR capacitance with illumination) gives:
\begin{equation}
U(t)=-E_0\times \ln \left[1-\left(1-\exp(-{U(0)\over
E_0})\right)\exp(-{t\over\tau})\right].
\end{equation}
Here it is seen that $\tau$, from Eq. (\ref{eq:Kin3}), is the
characteristic relaxation time of the photovoltage.
The photocathode is then illuminated by the second short  
laser pulse at time $t$ at which time the photovoltage is $U(t)$.
Using Eq. (\ref{PhVo}), the expression for the quantum yield $Y(t)$ 
of the probe pulse relative to the
yield with no photovoltage effect is then given by:
\begin{equation}
\frac{Y(t)}{Y_0}=1+{\frac{E_0}{\tilde \Delta}}\times
\ln \left[1-\left(1-\exp(-{U(0)\over
E_0})\right)\exp(-{t\over\tau})\right]. \label{eq:Pump}
\end{equation}
Fig. 6 shows the time variation
of this ratio for samples 1b, 2a, 2b, and 3. 
At large times $t$ it is seen that the photovoltage has 
decayed and the yield ratio asymptotically approaches one. 
The solid lines in Fig. 6 are the results of a fit to
Eq. (\ref{eq:Pump}).

The data of Fig. 2 for $Y(F)$ vs $\sqrt{F}$ are used to determine 
the quantity $\delta U(F)/\Delta$ using Eqs. (\ref{Aten}) and (\ref{bias}).
With  $\delta U = 47$ meV for a bias voltage of $- 120$ kV, the average
zero-field NEA value is determined to be $\Delta$ = 133 meV 
($\tilde{\Delta}=\Delta + \delta U$ = 180 meV). 
The data of Fig. 4 for $J\cdot Y(J)$ vs. $J$ are used to determine
the quantities $E_0/\tilde{\Delta}$ and $j_{p,0}$ using Eq. (\ref{eq:Yfin}) 
from which the parameter $E_0$ is calculated. 
The data of Fig. 5 for $\tau^\prime$ vs. $J$ are used to determine 
the quantity $\tau$ and independently determine $j_{p,0}$ using Eq. (\ref{eq:tau}). 
Finally, the pump-probe data of Fig. 6 are used as an independent method 
to determine the relaxation time $\tau$ using Eq. (\ref{eq:Pump}).
The average values of these parameters resulting from the fits are tabulated
in Table I. 

\section{Discussion}
The most important and theoretically predictable parameter is $E_0$.
The fit procedure yields $E_0 = 45$ meV for the average of samples 1a  
and 1b, and 62 meV for the average of samples 2a and 2b.
These values are consistent with the values expected from the light
hole contribution to the restoring current given by
Eq. (\ref{eq:Eoo}): $E_0 = 43$ meV for
sample 1, and  $E_0 = 61$ meV for sample 2.

The experimental results for the quantity $j_{p,0}$ are tabulated in
Table 1. These values are about 100 times smaller than  predicted by 
Eq. (\ref{eq:j0}), which suggests that the hole capture cross section
of the surface Cs states is much smaller than the value expected from
the  Schottky barrier model with a metallic Cs layer. Also, the
parameter $j_{p,0}$ is exponentially dependent on the initial 
surface band bending $V$, which varies in the range 0.3-0.5 eV depending
on the activation details, adding to the uncertainty. 

The observation of the linear dependence of $B_N$ on the
NEA value $\Delta$ predicted by Eq. (\ref{A7})
is strongly supported by the data on the variation of
$Y$ with the bias field. Numerical values for $B_N$ were calculated from the 
sample yield measurements using Eq. (\ref{eq:Y}) and were found 
to be in the range $0.09 \leq B_N \leq 0.26$.  
These values, together with Eq. (\ref{A7}) and the value 
for the optical phonon energy $\hbar\omega_0$ = 36 meV for GaAs,
are used to evaluate the electron emission rate  
$\langle \nu_{\rm emi}\rangle$ in vacuum. 
The value of $\tau_0$ in Eq. (\ref{A6}) was estimated for GaAs
from quantum well magneto-luminescence experiments to be 
0.05 ps\cite{MRP}, which gives $\langle\nu_{\rm emi}\rangle
\approx 0.7\ {\rm ps}^{-1}$.
This value is close to the values found
experimentally in the time-resolved electron emission
measurements from thin GaAs layers \cite{Schu}.
 

\section{Conclusions}
The present experimental results show that the surface escape
probability from NEA GaAs surfaces is in the range of
$B_N \le 0.26$ and is a linear function of the NEA energy 
including photovoltage and bias voltage effects. 
For well activated photocathodes, the quantum yield is observed to
increase linearly with the square-root of the bias field.  
The NEA value $\Delta$ is determined from the slope of this 
dependence giving values of $\Delta$ which are similar for all
well activated samples. 

The time evolution of
the photovoltage and the photovoltage dependence on excitation laser
intensity have been studied for thin GaAs samples with varying doping
concentrations using a long laser pulse technique and a second 
technique using short pulse lasers in a pump-probe configuration.  
The results of these measurements show that the 
photovoltage effect has a strong doping concentration dependence
decreasing with increased doping, and diminishing to zero at
a doping level of 5$\times$10$^{19}$cm$^{-3}$. 
The experimental determinations of the parameter $j_{p,0}$ are about 
1\% of the values expected from a Schottky barrier model, indicating that
the surface Cs layer has a non-metallic nature. 

The overall experimental results are consistent with a model based on 
electron energy relaxation by multiple phonon emission in the band 
bending region and tunneling through the surface barrier to vacuum.

\subsection*{Acknowledgments}
A.V.S. thankfully acknowledges the hospitality of SLAC
during his stay in USA. We thank R. Alley and T. Galetto for technical 
assistance. This work was supported by
Department of Energy contract  DE--AC03--76SF00515(SLAC) and 
DE-FG02-95ER40896(Wisconsin).
\pagebreak

\begin{appendix} 
{\bf APPENDIX Electron Kinetics in the BBR: Surface Escape Probability}

In this appendix an expression for the surface escape probability $B_N$ 
is derived. For unstrained GaAs photocathodes the EDC curves are spread 
over a broad energy band with a width close to the NEA value $\Delta$,
implying a long electron stay in the BBR accompanied by energy relaxation.
In both bulk and quantum well GaAs structures the most effective mechanism 
for energy loss is optical phonon emission. One can assume that this 
mechanism also dominates in the BBR. 
 
In this case the fraction of energy lost in one scattering event
is much smaller than $\Delta$. It is then possible to describe the 
energy relaxation and emission into vacuum by a Fockker-Planck 
\cite{LifPit,API} type
equation for the electron density $n(\epsilon)$ in the BBR at a given
energy $\epsilon$:
\begin{equation}
\frac{\partial n(\epsilon)} {\partial t} =- \frac{\partial}{\partial\epsilon
}\left[\frac{\epsilon}{\tau_{\epsilon}}n(\epsilon)-D(\epsilon)\frac{\partial
n(\epsilon)}{\partial\epsilon}\right]- \nu_{emi}(\epsilon)n(\epsilon) =0,
\eqnum{A1}\label{A1}
\end{equation}
where $\epsilon$ is the electron energy measured 
downward from the conduction band edge. 
The first term in square brackets describes the flow of electrons through
the states with energy $\epsilon$ due to phonon emission and the second term 
corresponds to diffusion in energy space. In this equation $\tau_{\epsilon}$ is the 
energy relaxation time, $\nu_{emi}(\epsilon)$ is the rate of electron
emission into vacuum, and $D(\epsilon)$ is the diffusion coefficient.   
The emission current density through the energy band $0 \leq \epsilon \leq
\Delta$ can be calculated as:
\begin{equation}
j_{emi} = q\int^{\Delta}_0 \nu_{emi}(\epsilon)n(\epsilon)d\epsilon . 
\eqnum{A2}\label{A2}
\end{equation}

To solve Eq. (\ref{A1}) we note that the diffusion 
term, for a broad energy distribution $\Delta
\gg kT$ in the electron flow, is of order $kT/\Delta \ll 1$ and can
be neglected. 
Then Eq. (\ref{A1}) reduces to:
\begin{equation}
\frac{\partial}{\partial\epsilon}\left[\frac{\epsilon}{\tau_{\epsilon}}n(\epsilon)\right] = 
\nu_{emi}(\epsilon)n(\epsilon). \eqnum{A3}\label{A3}
\end{equation} 
The calculation of the ratio of the emission current through
the energy band $0 \le \epsilon \le \Delta$ to the electron flow at
$\epsilon=0$ using Eq. (\ref{A2}) and (\ref{A3}) results in:
\begin{equation}
B_N= 1 - \exp \left[
- \int_0^{\Delta}\frac{\nu_{\rm emi}(\epsilon)\tau_{\epsilon}(\epsilon)}{\epsilon}
d\epsilon \right]. \eqnum{A4}\label{A4}
\end{equation}
The rate of electron emission in vacuum is controlled by the transparency of the
thin atomic-width barrier at the surface. The transparency
is presumably a slowly decreasing function of the electron energy
$\epsilon$, and therefore can be replaced in Eq. (\ref{A4})
by its average value $\langle \nu_{\rm emi}\rangle$. The surface escape probability
for the case $B_N \ll 1$ can then be written as:
\begin{equation}
B_N= \langle \nu_{\rm emi}\rangle \tau_d, \ \
\tau_d = \int_0^{\Delta} \tau_\epsilon {d\epsilon\over \epsilon}.
\eqnum{A5}\label{A5}
\end{equation} 

For the case where the dominant energy relaxation mechanism is optical
phonon emission, the following relation is a valid approximation\cite{API}:
\footnote{$\tau_0$ is assumed to be independent of $\epsilon$, a 
good approximation for the conditions of this experiment.}
\begin{equation}
\frac{\tau_{\epsilon}(\epsilon)}{\epsilon} = \frac{\tau_0}{\hbar\omega_0}, 
\eqnum{A6}\label{A6}
\end{equation}
where $\hbar\omega_0$ is the optical phonon energy and $\tau_0$ is the 
characteristic time for phonon emission. 
With these assumptions, $\tau_d$ and $B_N$ are linear functions of the NEA
value $\Delta$, giving:
\begin{equation}
B_N = \frac{\Delta}{E_B},\ 
E_B = \frac{\hbar\omega_0}{\langle\nu_{\rm emi}\rangle\tau_0}.
\eqnum{A7}\label{A7}
\end{equation}

The linear dependence of the surface escape probability on $\Delta$ 
can be obtained from a more general analysis of 
electron energy diffusion in the band bending region and does 
not depend on a surface density of states. This result,
the linear dependence of $B_N$ on $\Delta$,  is therefore not altered 
by band bending variations resulting from optical pumping. 

\end{appendix}

\newpage
\oddsidemargin=0cm

\pagebreak

\begin{table}[p]
\def\listtablename{    }

\hspace{0.8cm}

\begin{tabular}{|ccc|c|c|c|c|c|c|}
\multicolumn{3}{|c|}{Samples}&1a&1b&2a&2b&3&4\\ \hline
&p & 10$^{19}$ cm$^{-3}$&0.5&0.5&1&1&2&5\\
&$<Y>$& \%&0.6&0.45&0.9&0.3&0.4&0.4\\
&$\tau$ & ns& 160&130&78&63& 6& $<$ 4\\
&$E_0$& meV& 45&46&58&66&-&-\\
&$ j_{p,0}$/q& 10$^{18}$cm$^{-2}$s$^{-1}$&1.3&1.2&3.1&3.8&-&-\\ 
\end{tabular}
\vspace{0.25in}
\caption{The parameters of differently doped GaAs cathode layers
as determined from the high intensity excitation experimental data.
Here $<Y>$ is the surface averaged quantum yield as determined
from low bias voltage measurements. samples 1a, 1b and 2a, 2b
are different activations for samples 1 and 2 respectively.} 
\label{Tab1}
\end{table}
\pagebreak
\centerline{\bf LIST OF FIGURES}
\vspace{0.25in}
\begin{enumerate}
\item
Energy band diagram of a GaAs
active layer showing the  negative electron 
affinity vacuum level for various conditions. 
(a) The NEA variation with the bias field $\delta U(F)$
and (b) the photovoltage $U(J)$ with the arrows indicating the direction
of the movement of the NEA level. The photoemission and hole recombination
processes are shown schematically in Fig. 1b.
\item
The bias field dependence of the quantum yield for
various photocathodes.  Solid circles: sample 1a; Open circles: sample
3; Solid squares: zinc-doped strained GaAs; Open squares: carbon-doped
1$\mu$m-thick GaAs. The quantum yields have been rescaled as 
indicated in the figure. 
\item
The temporal profiles of the electron emission current using a long
laser excitation pulse for
(a) sample 1b, (b) sample 2a, and (c) sample 3. The laser intensity is
varied from 1 W/cm$^2$ to 150 W/cm$^2$.
\item
The steady state emission current 
as a function of the light excitation intensity for (a) sample 1a and 1b,
(b) sample 2a and 2b, and (c) sample 3. The solid lines are fitted results
to Eq. (\ref{eq:Yfin}).
\item
The saturation time constant $\tau^{\prime}$ of the photovoltage 
as a function of excitation light intensity for
(a) sample 1a and 1b, and (b) sample 2a and 2b.
The solid lines are fitted results to Eq. (\ref{eq:tau}).
\item
The quantum yield as a function of delay time between the pump
and probe pulses from pump-probe measurements for samples 1b, 2a, 2b, and 3. 
The solid lines are fitted results to Eq. (\ref{eq:Pump}).
\end{enumerate}
\pagebreak
\centerline{\epsfig{file=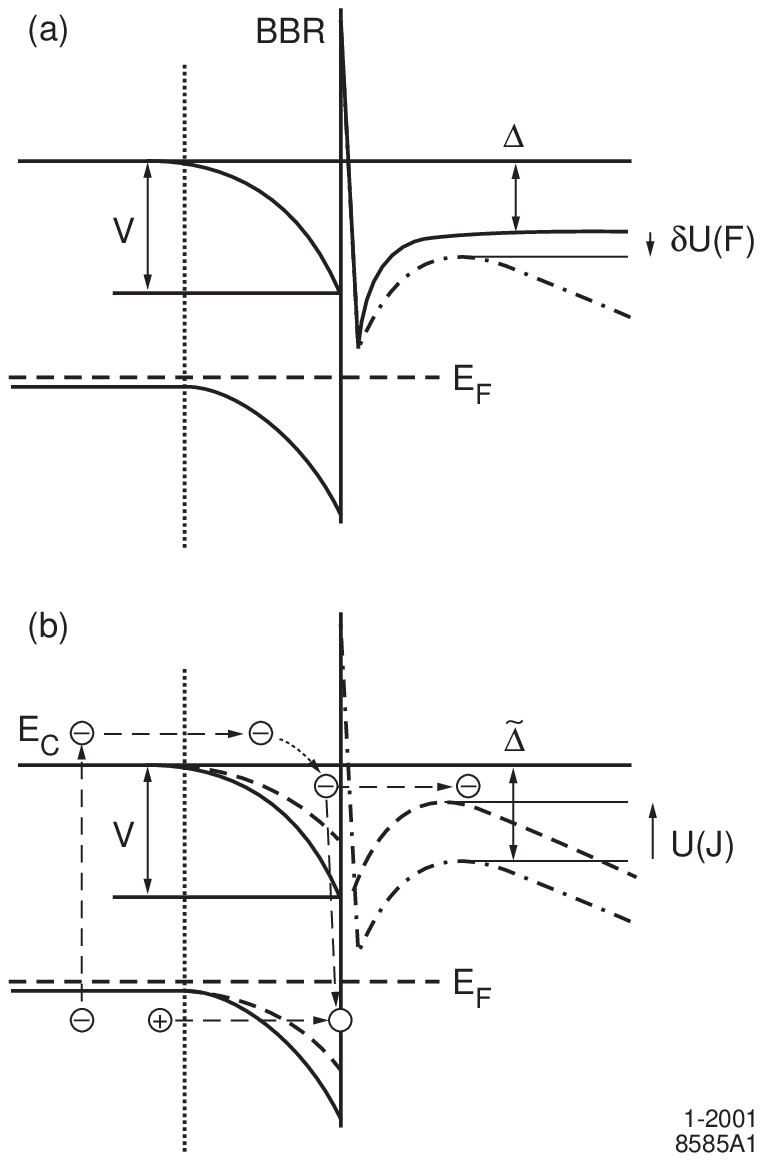}}
\pagebreak
\centerline{\epsfig{file=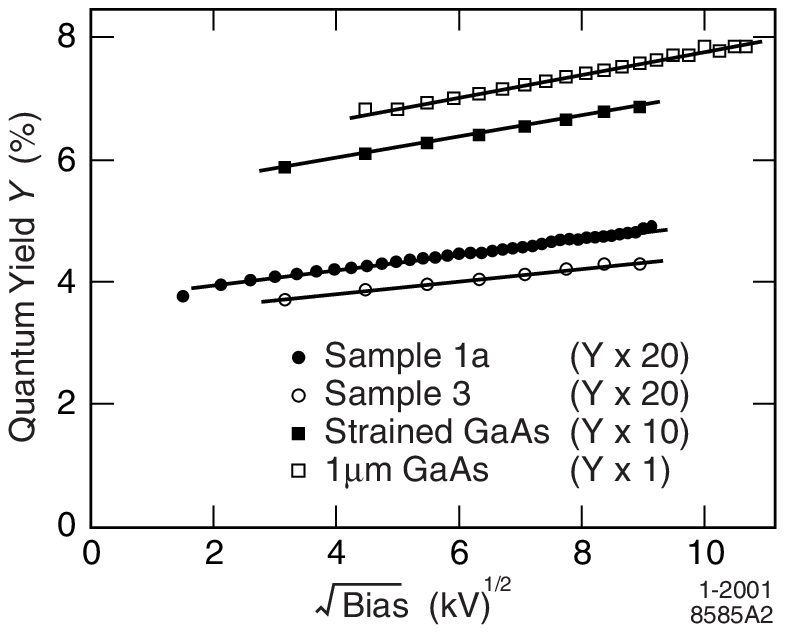}}
\pagebreak
\centerline{\epsfig{file=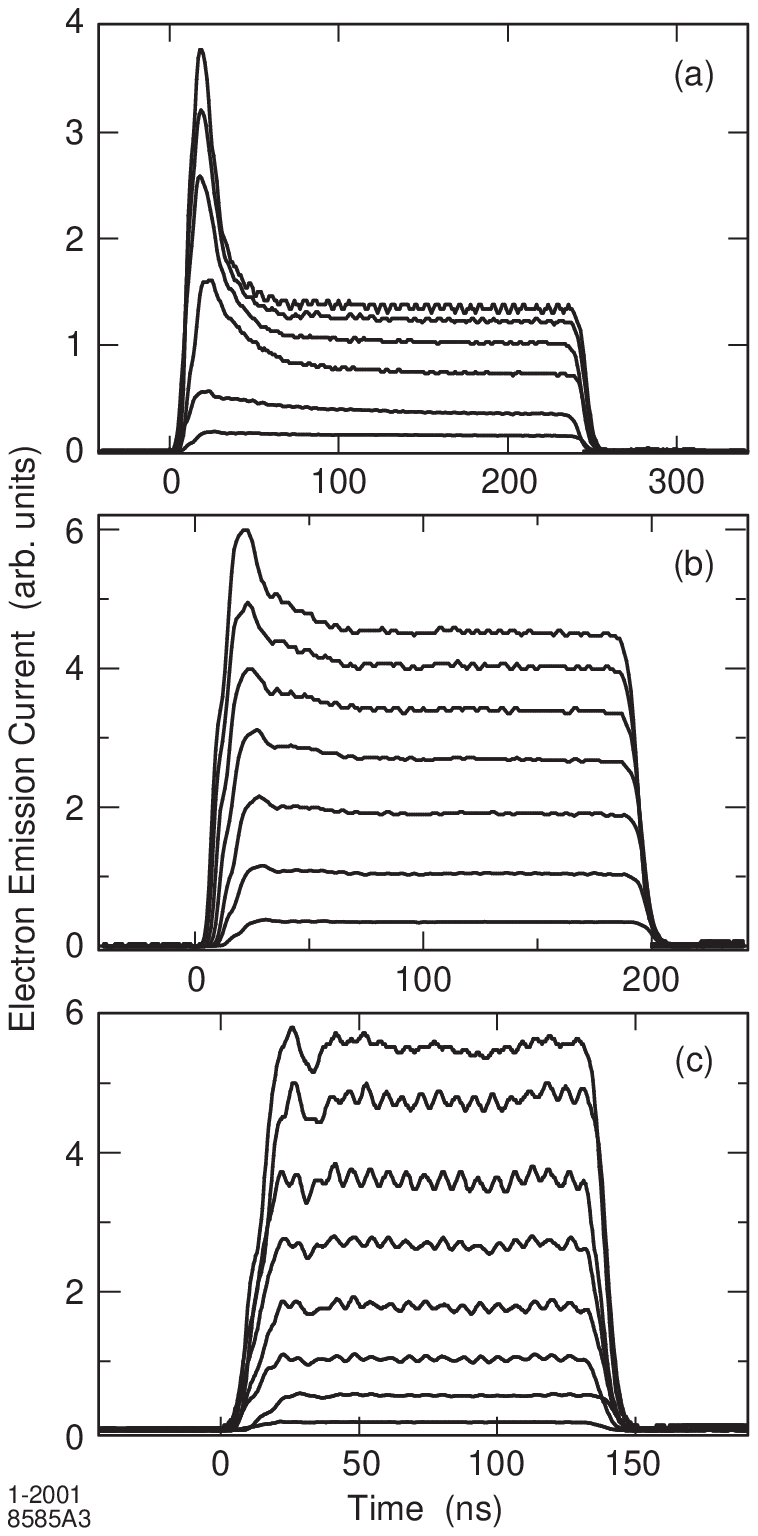}}
\pagebreak
\centerline{\epsfig{file=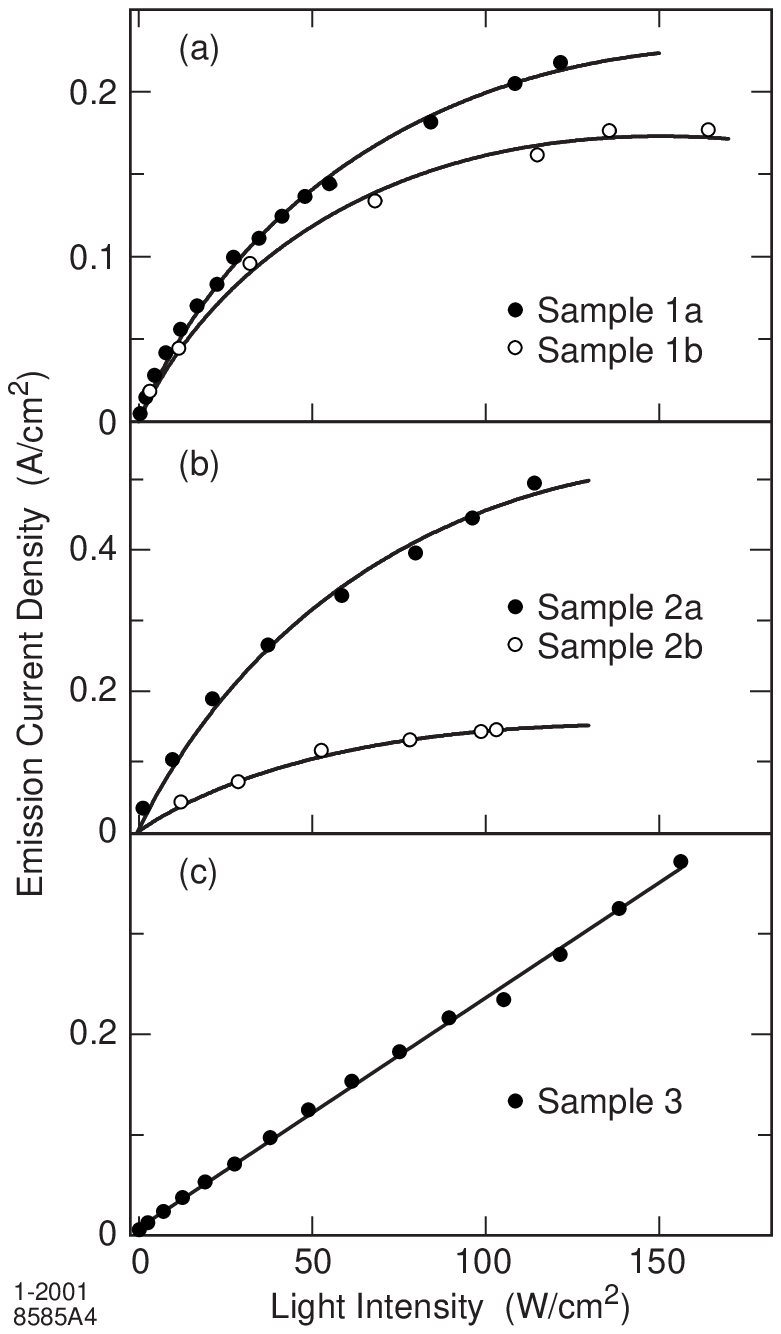}}
\pagebreak
\centerline{\epsfig{file=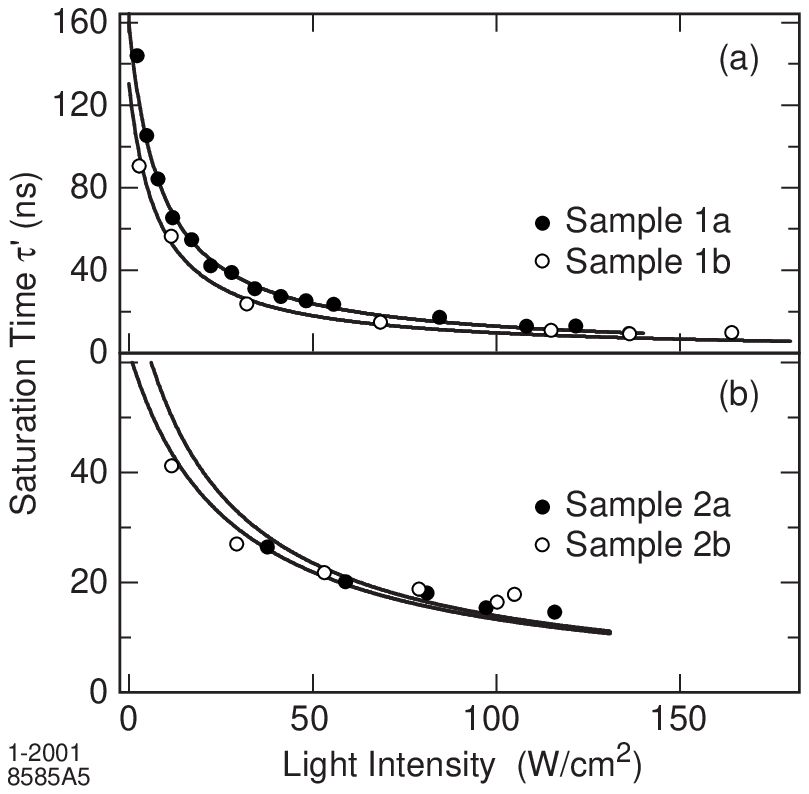}}
\pagebreak
\centerline{\epsfig{file=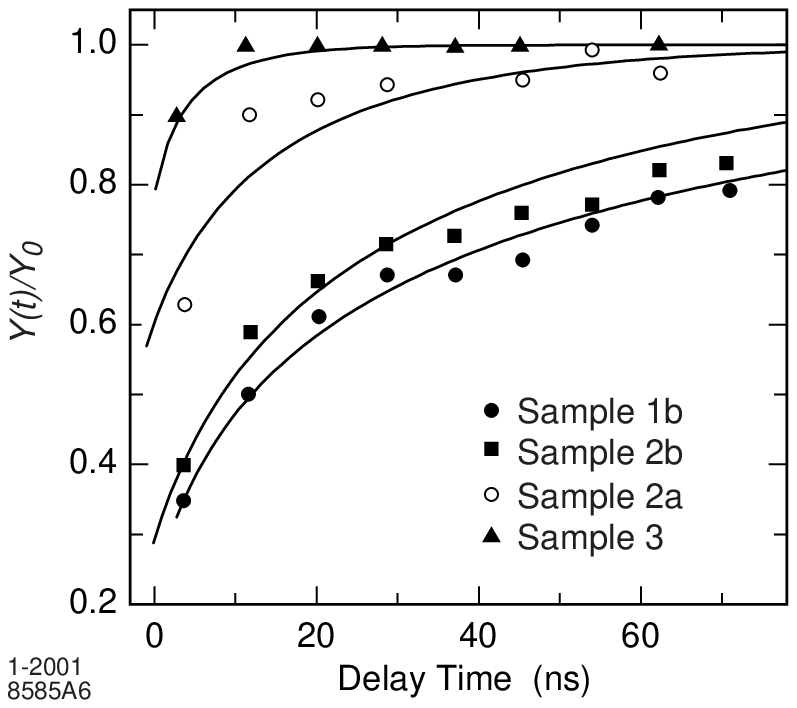}}
\end{document}